\documentclass[journal,twoside,web]{ieeecolor}
\usepackage{tmi}
\usepackage{cite}
\usepackage{amsmath,amssymb,amsfonts}
\usepackage{algorithmic}
\usepackage{graphicx}
\usepackage{textcomp}
\usepackage{multirow}
\usepackage{xcolor}
\usepackage{booktabs}
\usepackage{makecell}
\usepackage{hyperref}
\usepackage{xcolor}
\usepackage[ruled]{algorithm2e}
\usepackage{changes}
\usepackage{bm}
\def\BibTeX{{\rm B\kern-.05em{\sc i\kern-.025em b}\kern-.08em
  T\kern-.1667em\lower.7ex\hbox{E}\kern-.125emX}}
\begin{document}
\title{DoseDiff: Distance-aware Diffusion Model for Dose Prediction in Radiotherapy}
\author{Yiwen Zhang, Chuanpu Li, Liming Zhong, Zeli Chen, Wei Yang, and Xuetao Wang
\thanks{This work was supported in part by the National Natural Science Foundation of China under Grant 82172020 and Grant 62101239, in part by the Guangdong Basic and Applied Basic Research Foundation under Grant 2022A1515011336, and in part by the Guangdong Provincial Key Laboratory of Medical Image Processing under Grant 2020B1212060039. \emph{(Yiwen Zhang and Chuanpu Li are co-first authors.)(Corresponding authors: Xuetao Wang; Wei Yang)}}
\thanks{Yiwen Zhang, Chuanpu Li, Liming Zhong, Zeli Chen, and Wei Yang are with the School of Biomedical Engineering, Southern Medical University, Guangzhou 510515, China, and also with the Guangdong Provincial Key Laboratory of Medical Image Processing, Guangzhou 510515, China (e-mail: whisneyzyw@gmail.com; lichuanpugm@gmail.com; limingzhongmindy@gmail.com; dongyuczl@gmail.com; weiyanggm@gmail.com).}
\thanks{Xuetao Wang is with the Department of Radiation Therapy, The Second Affiliated Hospital, Guangzhou University of Chinese Medicine, Guangzhou 510006, China (e-mail: wangxuetao0625@126.com).}}

\maketitle

\begin{abstract}
Treatment planning, which is a critical component of the radiotherapy workflow, is typically carried out by a medical physicist in a time-consuming trial-and-error manner. Previous studies have proposed knowledge-based or deep-learning-based methods for predicting dose distribution maps to assist medical physicists in improving the efficiency of treatment planning. However, these dose prediction methods usually fail to effectively utilize distance information between surrounding tissues and targets or organs-at-risk (OARs). Moreover, they are poor at maintaining the distribution characteristics of ray paths in the predicted dose distribution maps, resulting in a loss of valuable information. In this paper, we propose a distance-aware diffusion model (DoseDiff) for precise prediction of dose distribution. We define dose prediction as a sequence of denoising steps, wherein the predicted dose distribution map is generated with the conditions of the computed tomography (CT) image and signed distance maps (SDMs). The SDMs are obtained by distance transformation from the masks of targets or OARs, which provide the distance from each pixel in the image to the outline of the targets or OARs. We further propose a multi-encoder and multi-scale fusion network (MMFNet) that incorporates multi-scale and transformer-based fusion modules to enhance information fusion between the CT image and SDMs at the feature level. We evaluate our model on two in-house datasets and a public dataset, respectively. The results demonstrate that our DoseDiff method outperforms state-of-the-art dose prediction methods in terms of both quantitative performance and visual quality.
\end{abstract}


\section{Introduction}
\label{sec:introduction}

\IEEEPARstart{R}{adiation} therapy (RT) is an essential cancer treatment modality; approximately 50\% of cancer patients receive RT during their course of illness and it contributes to around 40\% of curative treatment cases \cite{sahiner2019deep,delaney2005role}. Treatment planning plays an important part in the current RT workflow as it is used to determine the optimal radiation dose, technique, and schedule to target cancer while minimizing exposure to healthy tissue, with the aim of maximizing effectiveness and minimizing side effects. Clinically, a medical physicist typically spends hours adjusting a set of hyper-parameters and weightings in a trial-and-error manner to ensure that the RT plan can achieve the desired treatment effect, which is time-consuming and labor-intensive \cite{craft2012improved,schreiner2011quality}. If an appropriate dose distribution map can be obtained in advance, the medical physicist will be able to use it as a reference, enabling the RT planning to be completed with fewer hyper-parameter adjustments \cite{zhan2022multi,song2020dose}. Therefore, dose prediction is of great value in enhancing efficiency and streamlining the workflow of RT.

Various approaches have been proposed for dose prediction in RT. Knowledge-based planning (KBP) provides a traditional paradigm for dose prediction that leverages planning information from historical patients to predict dosimetry for a new patient \cite{shiraishi2015knowledge,nwankwo2015knowledge}. Initially, some handcrafted features related to dosimetry are directly used to match historical patients with new patients, e.g., the overlap volume histogram \cite{wu2011data}, distance-to-target histogram \cite{song2015patient}, and planning volume shapes \cite{deshpande2016knowledge}. With the development of machine learning technologies, support vector regression, random forests, and gradient boosting have been used to find more effective features \cite{valdes2017clinical,mcintosh2015contextual,morin2018deep}. However, these KBP-based methods typically concentrate on predicting dosimetric endpoints or dose-volume histograms (DVHs), which are dosimetric-related statistical indicators and lack spatial information \cite{zhan2022multi}. Furthermore, the accuracy of methods that rely on traditional feature extraction often falls short of expectations.

Deep learning is currently a popular approach for dose prediction \cite{zhan2022multi,song2020dose,dong2020deep,fan2019automatic,kearney2018dosenet,kearney2020dosegan,kontaxis2020deepdose,nguyen20193d,nguyen2019feasibility,sumida2020convolution,yue2022dose,hu2021incorporating}. With its powerful feature extraction and analysis capabilities, a deep learning model can learn mapping relationships from computed tomography (CT) images and region-of-interest (ROI) masks, i.e., targets and organs-at-risk (OARs), to dose distribution maps. Nguyen et al. \cite{nguyen2019feasibility} and Fan et al. \cite{fan2019automatic} used a two-dimensional (2D) convolutional neural network (CNN) to predict dose distribution slice by slice, with the ultimate objective of achieving three-dimensional (3D) dose prediction. To make full use of spatial information, Kearney et al. \cite{kearney2018dosenet} proposed DoseNet, a modified 3D UNet, for volumetric dose prediction. Recent research has focused on constructing more effective loss functions to improve the accuracy of dose prediction, e.g., adversarial loss \cite{kearney2020dosegan} and ROI-weighted loss \cite{zhan2022multi}.

Most deep-learning-based dose prediction methods use CT images and ROI masks directly as model inputs; however, they rarely exploit distance information between surrounding tissues and targets or OARs. In RT planning, distance information is important as the relative positions of ROIs determine their mutual influence, e.g., the dose levels of  tissues close to targets are relatively large, whereas those of the tissues near OARs are relatively small \cite{yue2022dose,yuan2012quantitative}. Although CNNs are capable of implicitly learning distance relationships among ROIs from masks, most of deep-learning-based methods use a slice-based or patch-based training strategy owing to memory constraints, which means they are unable to leverage useful information from mask images when the slice or patch contains little or no ROI. While some methods incorporate attention modules into the networks to perceive long-dependency range information, the attention perception of these modules remains confined to slices or blocks \cite{jiao2023transdose,xiao2022transdose}. To address the aforementioned challenge, certain studies have introduced distance information into neural networks through distance maps. For example, Kontaxis et al. \cite{kontaxis2020deepdose} fed the distance map from the central beamline to the model to improve the accuracy of dose prediction. However, their method requires knowledge of the parameters of beamlets, which is often not practical in a clinical setting. Zhang et al. \cite{zhang2020predicting} introduced a distance image of planning target volumes (PTV) for esophageal radiotherapy dose prediction, but ignored OARs and other targets. Yue et al. \cite{yue2022dose} utilized ROI masks for distance map calculation and proposed a normalization technique to rescale the numerical range of the distance map for network training, but their distance map only considered pixel distance in image space, rather than real-world physical distance, and had a limited dynamic range. In addition, these methods simply concatenated the distance maps and CT images as network inputs, restricting the potential performance of their model. Various studies have shown that feature-level fusion is more capable of capturing complex relationships among different modalities compared with input-level fusion \cite{zhou2019review,zhou2020hi,meng2023msmfn}. 

Ensuring the authenticity and feasibility of the predicted dose distribution map is also a challenge, e.g., it involves ensuring that the radiation paths in the map adhere to straight-line propagation and that the predicted cumulative dose distribution can be achieved within an acceptable number of treatment fields. However, most of previous deep-learning-based dose prediction methods only learn the mapping from input to output images and so cannot fully capture the prior data distribution of real dose distribution map. As a result, the predicted dose distribution maps appear excessively smooth and distort ray path characteristics. To address this problem, Wang et al. \cite{wang2022deep} proposed a beam-wise dose network to refine the predicted dose within beam masks, but these artificially predefined beam masks derived from the PTV masks are quite coarse. Some studies have adopted frameworks based on generative adversarial networks (GANs) to enhance the realism of the predicted dose distribution map \cite{zhan2022multi,nwankwo2015knowledge}, but GANs can be challenging to train and often drop modes in the output distribution \cite{gulrajani2017improved,metz2016unrolled,saharia2022palette}. Recently, diffusion models \cite{ho2020denoising,song2021scorebased} have attracted attention as powerful generative models that are capable of modeling complex image distribution. Diffusion models have been shown to provide superior image sampling quality and more stable training compared with GANs \cite{dhariwal2021diffusion}. Conditional diffusion models introduce conditional information to guide the model in generating specific images. Researchers have applied conditional diffusion models to various image-to-image translation tasks, e.g., image inpainting \cite{xie2023smartbrush}, colorization \cite{saharia2022palette}, super-resolution \cite{gao2023implicit,saharia2022image}, and medical image synthesis \cite{lyu2022conversion}, and achieved the SOTA results. Unlike the previous method of learning pixel mappings between input and output, conditional diffusion models can learn the data distribution from training images and sample the images that best match the given conditions based on the distribution. A conditional diffusion model also partitions image-to-image generation into a sequence of denoising steps, which typically recover the general outline initially and then produce details. This can be considered to be a recursive image generation method and has been proven to be effective by previous studies \cite{cai2019multi,ma2019coarse}. Despite these advantages of diffusion models, further exploration is required to fully realize their potential in the task of predicting dose distribution in RT. To the best of our knowledge, only Feng et al. \cite{feng2023diffdp} proposed a diffusion-based dose prediction (DiffDP) model for predicting the radiotherapy dose distribution of cancer patients. However, DiffDP is limited to utilizing 2D in-plane information for slice-based dose prediction. 

In this paper, we propose a novel distance-aware conditional diffusion model for dose prediction, named DoseDiff, which takes CT images and signed distance maps (SDMs) as conditions to accurately generate dose distribution maps. Based on the mechanism of the conditional diffusion model, we define dose prediction using a sequence of denoising steps, wherein the predicted dose distribution maps are generated from Gaussian noise images with the guidance of conditions. The SDMs are obtained by performing a distance transform on the masks to obtain indicates the distance of each voxel in the image from the contours of the ROIs in 3D space. Unlike Yue et al. \cite{yue2022dose}, who simply combined CT images and distance maps through channel concatenation at the input, we propose a multi-encoder and multi-scale Fusion Network (MMFNet) to enhance information fusion at the feature level. While the early fusion methods, such as channel concatenation, are straightforward and efficient, their performances are constrained by the presence of considerable noise and redundant information in the low-level features. Extensive experiments have demonstrated high-level feature fusion is valuable to the precise prediction in complex tasks \cite{wajid2021multimodal,hermessi2021multimodal}. We design independent encoders for CT images and SDMs, respectively, and perform information fusion on feature maps of multiple scales, which is able to integrate both low-level and high-level features. A fusion module based on self-attention mechanism \cite{vaswani2017attention,dosovitskiy2020image} is also proposed for inclusion in MMFNet to further fuse global information.

\begin{figure*}[!t]
\centerline{\includegraphics[width=1.\textwidth]{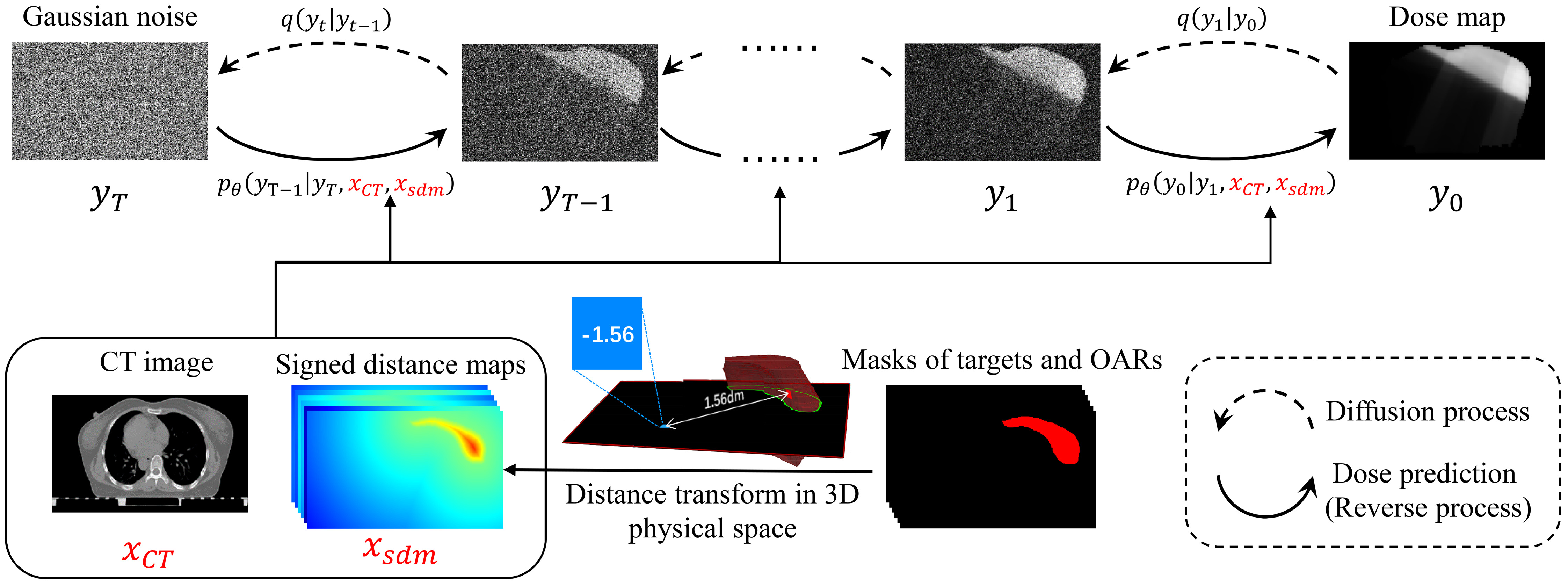}}
\setlength{\abovecaptionskip}{0.05cm}
\caption{The overall workflow for DoseDiff.}
\label{Fig1}
\end{figure*}

\section{Method}
\subsection{Background}
Conditional diffusion models represent a class of conditional generative methods aimed at transforming a Gaussian distribution to an empirical data distribution \cite{song2021scorebased,ho2020denoising,nichol2021improved}. This usually involves two processes in opposite directions: the forward (diffusion) process and reverse process. We denote by $x$ and $y_0$ the condition and target domain images, respectively. The forward process aims to collapse the target image distribution $p(y_0)$ to a standard Gaussian distribution $\mathcal{N}(\bf{0},\rm{\bf{I}})$ by gradually adding Gaussian noise of varying scales to the target image:
\begin{equation}
\begin{aligned}
q\left(y_t \mid y_{t-1}\right)=\mathcal{N}\left(y_t ; \sqrt{\alpha_t} y_{t-1},\left(1-\alpha_t\right) \rm{\bf{I}}\right);
\end{aligned}
\end{equation}
\begin{equation}
\begin{aligned}
q\left(y_t \mid y_0\right)=\mathcal{N}\left(y_t ; \sqrt{\bar{\alpha}_t} y_0,\left(1-\bar{\alpha}_t\right) \rm{\bf{I}}\right),
\end{aligned}
\end{equation}
where $t\in[1,T]$ is the timestep and $T$ denotes the total number of diffusion steps. The scale of the added Gaussian noise is determined by a set of hyper-parameters, ${\alpha}_t$, that change depending on $t$, with $\bar{\alpha}_t=\prod_{s=1}^t \alpha_s$. Using the theoretical assumption of the diffusion model, when $T$ is sufficiently large, $p(y_T)$ can be approximated as a standard Gaussian distribution \cite{ho2020denoising,song2021scorebased}.

The target image is generated by fitting other distributions that are parameterized by $\theta$ in the reverse process:
\begin{equation}
\label{eqs3}
\begin{aligned}
p_\theta\left(y_{t-1} \mid y_t\right)=\mathcal{N}\left(y_{t-1} ; \mu_\theta\left(y_t, x, t\right),\left(1-\alpha_t\right) \rm{\bf{I}}\right),
\end{aligned}
\end{equation}
where $\mu_\theta\left(y_t, x, t\right)$ can be parameterized by a neural network $\epsilon_\theta\left(y_t, x, t\right)$. The objective of training the neural network is to minimize the variational upper bound of the negative log-likelihood:
\begin{equation}
\label{eqs4}
\begin{aligned}
\mathcal{L}_\theta=\mathbb{E}_{y_0, x, t, \epsilon}\left\|\epsilon-\epsilon_\theta\left(y_t, x, t\right)\right\|,
\end{aligned}
\end{equation}
where $\epsilon \sim \mathcal{N}(0,\rm{\bf{I}})$.

\subsection{DoseDiff}
In contrast to previous methods, the conditional diffusion model can model the data distribution, which enables it to perceive the dose distribution characteristics in the real dose distribution map. Furthermore, the conditional diffusion model exhibits more stable training and yields higher image quality compared with GAN-based methods. Therefore, we propose DoseDiff for dose prediction, as shown in Fig. \ref{Fig1}. DoseDiff is essentially a conditional diffusion model conditioned by CT images and SDMs, which considers dose prediction as a sequence of denoising steps, i.e., the reverse process. To adapt the two conditional terms, we redefine the neural network as $\epsilon_\theta\left(y_t, x_{C T}, x_{s d m}, t\right)$ and accordingly modify the reverse process and objective function outlined in Eqs. \ref{eqs3} and \ref{eqs4}:
\begin{equation}
\begin{aligned}
p_\theta\left(y_{t-1} \mid y_t\right)=\mathcal{N}\left(y_t ; \mu_\theta\left(y_t, x_{C T}, x_{s d m}, t\right),\left(1-\alpha_t\right) \rm{\bf{I}}\right),
\end{aligned}
\end{equation}
\begin{equation}
\begin{aligned}
\mathcal{L}_\theta=\mathbb{E}_{y_0, x_{C T}, x_{s d m}, \epsilon}\left\|\epsilon-\epsilon_\theta\left(y_t, x_{C T}, x_{s d m}, t\right)\right\|,
\end{aligned}
\end{equation}
where $x_{CT}$ and $x_{sdm}$ are the CT image and SDM, respectively, and $\epsilon \sim \mathcal{N}(\bf{0},\rm{\bf{I}})$.

The conditional diffusion model assumes that the reverse process is a Markov chain, and diffusion step $T$ should be sufficiently large. Therefore, the inference of DoseDiff consumes significant computational time. To reduce the inference time, we adopt the accelerated generation technology of the denoising diffusion implicit model (DDIM) \cite{song2020denoising}. Based on the non-Markovian hypothesis, the DDIM allows a reduced number of sampling steps (fewer than $T$) in the inverse process. Furthermore, the implementation of DDIM requires modifications only to the sampling technique of the inverse process, with no impact on the training and forward processes. Let $\tau$ represent a sub-sequence of $[1,…,T]$ of length $S$ with $\tau_S=T$; then, the accelerated reverse process of DoseDiff can be expressed as:
\begin{equation}
\label{eqs7}
\begin{aligned}
&p_\theta\left(y_{\tau_{i-1}} \mid y_{\tau_i}\right)= \mathcal{N}\left(\sqrt{\bar{\alpha}_{\tau_{i-1}}} y_{0|\tau_i}+\sqrt{1-\bar{\alpha}_{\tau_{i-1}}-\sigma_{\tau_i}^2}\right. \\
& \left.\cdot \epsilon_\theta(y_{\tau_i}, x_{C T}, x_{s d m},\tau_i),\sigma_{\tau_i}^2 \rm{\bf{I}}\right) \text { if } i \in[S], i>1,
\end{aligned}
\end{equation}
\begin{equation}
\label{eqs8}
\begin{aligned}
p_\theta\left(y_0 \mid y_t\right)=\mathcal{N}\left(y_{0|t},\sigma_{t}^2 \rm{\bf{I}}\right) \text {otherwise},
\end{aligned}
\end{equation}
where $y_{0|t}=\left(y_t-\sqrt{1-\bar{\alpha}_t} \epsilon_\theta\left(y_t, x_{C T}, x_{s d m}, t\right)\right) / \sqrt{\bar{\alpha}_t}$ and $\sigma_t=\sqrt{(1-\bar{\alpha}_{t-1})/(1-\bar{\alpha}_t)}\sqrt{1-\bar{\alpha}_t/\bar{\alpha}_{t-1}}$.

\begin{figure*}[!t]
\centerline{\includegraphics[width=1.\textwidth]{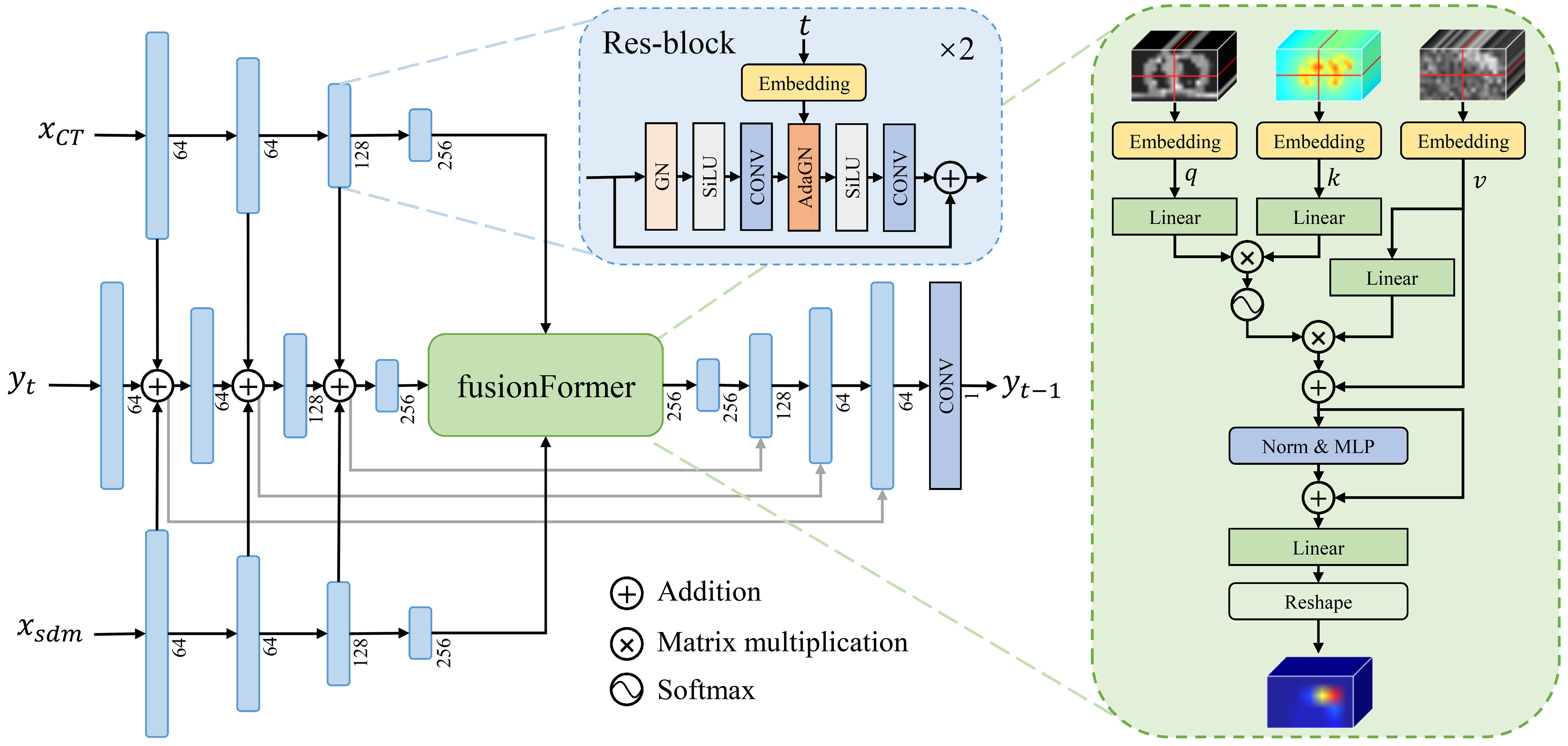}}
\setlength{\abovecaptionskip}{0.05cm}
\caption{Architecture of MMFNet for dose prediction.}
\label{Fig2}
\end{figure*}

\subsection{Signed distance map}
Most previous dose prediction methods have directly used masks to provide location and distance information related to ROIs. A limitation of using masks in this way is that current slice-based and patch-based CNN training strategies cannot guarantee that all foreground regions in the mask images are completely included in the sampled images. Moreover, although CNNs are capable of implicitly learning distance relationship among ROIs from mask images, their convolution kernels can only extract local features, and thus they may struggle to perceive long-range distance relationships. On the contrary, a distance map provides the distance from each voxel to the contour of the ROI in the 3D space of the image. The distance image value itself is capable of conveying distance information, regardless of incomplete sampling of the image or the limited receptive field of CNNs. The SDM uses sign to identify whether a voxel is inside or outside the ROI.

Let $M_c$ denote the mask image, where $c\in[1,C]$ and $C$ is the total number of targets and OARs. We denote by $M_c^{in}$, $M_c^{out}$, and $\partial M_{c}$ inside, outside, and boundary of the ROI, respectively. The SDM in image space (ISDM) is defined as:
\begin{equation}
\begin{aligned}
x_c^{i s d m}=\left\{\begin{array}{c}
\inf _{v \in \partial M_c}\mathcal{D}_i(u,v), u \in M_c^{i n} \\
0, u \in \partial M_c \\
-\inf _{v \in \partial M_c}\mathcal{D}_i(u,v), u \in M_c^{o u t}
\end{array}\right.,
\end{aligned}
\end{equation}
where $\mathcal{D}_i(u,v)=\sqrt{(i_u-i_v)^2+(j_u-j_v)^2+(k_u-k_v)^2}$ is the Euclidean distance between voxels $u$ and $v$ in image space, where $(i_u,j_u,k_u)$ and $(i_v,j_v,k_v)$ are voxel coordinates of $u$ and $v$. However, owing to inconsistent voxel resolution, ISDM values are not comparable between different images and have no practical connotation. Therefore, we consider spacing as a factor when calculating the SDM, transforming the ISDM from image space to physical space (PSDM): $\mathcal{D}_i(u,v) \rightarrow \mathcal{D}_p(u,v,\textbf{s})=\sqrt{i_\textbf{s}(i_u-i_v)^2+j_\textbf{s}(j_u-j_v)^2+k_\textbf{s}(k_u-k_v)^2}$, where $\textbf{s}(i_\textbf{s},j_\textbf{s},k_\textbf{s})$ is the voxel spacing of the image. Finally, the input SDM $x_{sdm}$ of the neural network $\epsilon_\theta\left(y_t, x_{C T}, x_{s d m}, t\right)$ is a collection of the SDMs of all ROIs:
\begin{equation}
\begin{aligned}
x_{sdm}=\left\{x_1^{psdm},x_2^{psdm},...,x_{C-1}^{psdm},x_C^{psdm}\right\}.
\end{aligned}
\end{equation}

\subsection{MMFNet}
The conditions of DoseDiff contain multimodal images, i.e., the CT image and SDM. To effectively extract and fuse their features, we propose MMFNet for $\epsilon _{\theta}$, as shown in Fig. \ref{Fig2}. MMFNet adopts three encoders used to extract the features of $x_{CT}$, $x_{sdm}$, and $y_t$, respectively. The encoder part accomplishes multi-scale information fusion by summing feature maps of the same resolution from the three inputs at every level. Our encoders and decoder are implemented in the same way as those of \cite{dhariwal2021diffusion}, with four down-sampling or up-sampling operations. Each level of encoders and decoder contains two residual blocks (Res-blocks), each of which is composed of one group normalization (GN) layer \cite{wu2018group}, two sigmoid linear units, two $3 \times 3$ convolution layers, one adaptive GN (AdaGN) layer \cite{dhariwal2021diffusion}, and one residual connection. Similarly, the timestep $t$ is embedded in each Res-block through AdaGN, which is defined as \cite{dhariwal2021diffusion}:

\begin{equation}
\begin{aligned}
\operatorname{AdaGN}\left(f, \mathbf{e_t}\right)=e_t^s \mathrm{GN}(f)+e_t^b,
\end{aligned}
\end{equation}

where $f$ is the input feature map and $\mathbf{e_t}=[e_t^s,e_t^b]$ is obtained from a linear projection of the timestep $t$. 

We propose a transformer-based fusion module, fusionFormer, for further global information fusion. Unlike the convolution layer used to extract local features, the transformer \cite{vaswani2017attention,dosovitskiy2020image} uses an attention mechanism to capture global context information. A currently popular approach to enhance model performance is to combine the strengths of CNNs and transformers for local and global feature extraction \cite{d2021convit,yuan2021incorporating}. To minimize memory cost, fusionFormer is added to the path at the lowest resolution level. In fusionFormer, the feature maps of $x_{CT}$, $x_{sdm}$, and $y_t$ are partitioned into $2 \times 2$ non-overlapping patches, i.e., the channels remain unchanged while evenly slicing along the length and height dimensions. The patch size is associated with the size of the original input image. The three feature maps are then transformed into three fundamental components of the attention using an embedded layer consisting of linear and normalization layers: the query ($f_q$), key ($f_k$), and value ($f_v$). As both $x_{CT}$ and $x_{sdm}$ are conditions, we use their feature maps as the query and key (or inverse) to calculate attention matrix. The attention matrix is then used to weight the feature map of the noisy dose distribution map, enabling conditional guidance. The attention block in the transformer can be expressed as follows:
\begin{equation}
\begin{aligned}
f_{a t t}=\operatorname{Softmax}\left(\beta\left(f_q\right) \times \gamma\left(f_k\right)^{\mathrm{T}}\right) \times \delta\left(f_v\right)+\delta\left(f_v\right),
\end{aligned}
\end{equation}
where $\beta(\cdot)$, $\gamma(\cdot)$, and $\delta(\cdot)$ are linear operations. Then, a feed-forward block is used for the nonlinear transformation of features:
\begin{equation}
\begin{aligned}
f_{f f b}=\operatorname{MLP}\left(\operatorname{LN}\left(f_{a t t}\right)\right)+f_{a t t},
\end{aligned}
\end{equation}
where $\operatorname{MLP}(\cdot)$ is a multi-layer perceptron consisting of two linear layers and a nonlinear activation function and $\operatorname{LN}(\cdot)$ is layer normalization. Last, a linear layer is used to adjust the dimensionality of the feature vector to the original size; it is then reshaped to be the same shape as the input feature maps.

\subsection{Implementation details}
Our network implementation was based on PyTorch, and the code was run on a server with two RTX 2080Ti GPUs. Following \cite{ho2020denoising}, we set the total number of diffusion steps ($T$) to 1000 and the forward process variances $\alpha_t$ to constants decreasing linearly from $\alpha_1 = 0.9999$ to $\alpha_T = 0.08$. To ensure the input value of the model fell within a suitable range, we set the distance unit for the PSDM to decimeters. We used the AdamW \cite{loshchilov2017decoupled} optimizer with a step-decay learning rate to train our model for one million iterations. The initial learning rate and batch size were 0.0001 and 8, respectively. Some simple online data augmentation operators, including random flipping, rotating, and zooming, were used on the training set to improve the generalization capacity of the model. To reduce the inference time, we set the reduced generation step $S$ in DDIM to 8. The full implementation is available at \url{https://github.com/whisney/DoseDiff}.

\section{Experiments}
\subsection{Datasets and Preprocessing}
\subsubsection{In-house Datasets}
The breast cancer and nasopharyngeal cancer (NPC) datasets were constructed by collecting data for patients receiving RT at Guangdong Provincial Hospital of Traditional Chinese Medicine, China, from 2016 to 2020. The patients in the breast cancer dataset underwent postoperative RT following breast-conserving surgery, whereas the patients in the NPC dataset received RT for primary lesions. The CT volumes were obtained using a Siemens Sensation Open (Siemens Healthcare, Forchheim, Germany) scanner. RT plans were received from a treatment-planning system (Philips, Pinnacle3, Netherlands). All the targets and OARs were delineated by experienced oncologists and all the plans were clinically approved. The details of the breast cancer and NPC datasets were as follows.

\textbf{\textit{Breast cancer dataset:}} The breast cancer dataset consisted of data from 119 patients. The in-plane pixel spacings of the CT images ranged from 0.77 to 0.97 mm with an average of 0.95 mm, and slice thicknesses were all 5.0 mm. The in-plane resolutions were $512 \times 512$, and the number of slices ranged from 50 to 129. We used a binary body mask to exclude unnecessary background from each image, and all the images were cropped by the minimal external cube of the body masks and then resized to $320 \times 192$ in-plane. The target areas of tumor bed (TB) and clinical target volume (CTV), and the OARs for heart, spinal cord, and left and right lungs were included in our experiments.

\textbf{\textit{NPC dataset:}} The NPC dataset consisted of data from 139 patients. The in-plane pixel spacings of the CT images ranged from 0.75 to 0.97 mm with an average of 0.95 mm, and slice thicknesses were 3.0 mm. The in-plane resolutions were $512 \times 512$, and the number of slices ranged from 69 to 176. All images were also masked and cropped by body masks but then resized to $448 \times 224$ in-plane. The target areas included gross tumor volume (GTV) and PTV, and the OARs included eyes, optic nerve, temporal lobe, brain stem, parotid gland, mandible, and spinal cord.

The intensity ranges for CT images and dose distribution maps were set to $[-1000,1500]$ HU and $[0,75]$ Gy, respectively. Both were uniformly and linearly normalized to $[-1,1]$ for training. Both datasets were split into training, validation, and test sets using a patient-wise ratio of $7:1:2$; these sets were used for model training, model selection, and performance evaluation, respectively.

\subsubsection{Public Dataset}
We utilize the public Head and neck cancer dataset from the AAPM \textit{OpenKBP} challenge \cite{babier2021openkbp}, which contains 200 training cases, 40 validation cases, and 100 testing cases. The RT plans were prescribed 70,  63, and 56 Gy in 35 fractions to the gross disease ($\mathrm{PTV}_{70}$), intermediate-risk target volumes ($\mathrm{PTV}_{63}$), and elective target volumes ($\mathrm{PTV}_{56}$). The spacing and size of all images were consistent at 3.906 mm $\times$ 3.906 mm $\times$ 2.5 mm and $128 \times 128 \times 128$, respectively. The target areas included PTV, brain stem, spinal cord, parotid gland, larynx, esophagus, and mandible.

\begin{table*}[!t]
\centering
\caption{Results of ablation studies for DoseDiff on the breast cancer dataset.}
\label{table1}
\begin{tabular*}{\textwidth}{@{\extracolsep{\fill}}cccccccccc}
\toprule
UNet & Diffusion model & PSDM & MS & FF & MAE (Gy)$\downarrow$ & SSIM$\uparrow$ & PSNR (dB)$\uparrow$ & Dose score (Gy)$\downarrow$ & Volume score (\%)$\downarrow$ \\
\midrule
$\checkmark$ & ~ & ~ & ~ & ~ & 3.416$\pm$1.741 & 0.797$\pm$0.077 & 19.675$\pm$4.175 & 11.206$\pm$7.456 & 18.910$\pm$9.909 \\
$\checkmark$ & ~ & ~ & $\checkmark$ & ~ & 2.863$\pm$1.199 & 0.811$\pm$0.059 & 20.900$\pm$3.570 & 9.228$\pm$5.114 & 15.039$\pm$8.963 \\
$\checkmark$ & ~ & ~ & ~ & $\checkmark$ & 2.984$\pm$1.392 & 0.813$\pm$0059 & 20.539$\pm$3.680 & 8.369$\pm$5.129 & 15.981$\pm$8.909 \\
$\checkmark$ & ~ & $\checkmark$ & ~ & ~ & 1.438$\pm$0.224 & 0.896$\pm$0.032 & 26.662$\pm$1.240 & 1.880$\pm$0.611 & 11.301$\pm$8.167 \\
$\checkmark$ & ~ & $\checkmark$ & $\checkmark$ & ~ & 1.420$\pm$0.211 & 0.890$\pm$0.034 & 26.710$\pm$1.140 & 1.150$\pm$0.333 & 1.718$\pm$1.357 \\
$\checkmark$ & ~ & $\checkmark$ & $\checkmark$ & $\checkmark$ & \textbf{1.221$\pm$0.241} & \textbf{0.913$\pm$0.030} & \textbf{27.813$\pm$1.649} & \textbf{1.047$\pm$0.438} & \textbf{1.256$\pm$0.740} \\
\midrule
~ & $\checkmark$ & ~ & ~ & ~ & 2.937$\pm$1.132 & 0.807$\pm$0.052 & 20.270$\pm$3.134 & 10.391$\pm$6.376 & 19.413$\pm$7.791 \\
~ & $\checkmark$ & ~ & $\checkmark$ & ~ & 2.654$\pm$1.156 & 0.820$\pm$0.054 & 21.333$\pm$3.475 & 8.317$\pm$5.361 & 14.790$\pm$9.077 \\
~ & $\checkmark$ & ~ & ~ & $\checkmark$ & 2.493$\pm$1.011 & 0.819$\pm$0.044 & 22.114$\pm$3.039 & 7.047$\pm$6.557 & 13.143$\pm$8.746 \\
~ & $\checkmark$ & $\checkmark$ & ~ & ~ & 1.223$\pm$0.206 & 0.903$\pm$0.209 & 27.710$\pm$1.539 & 1.575$\pm$0.440 & 7.031$\pm$7.549 \\
~ & $\checkmark$ & $\checkmark$ & $\checkmark$ & ~ & 1.201$\pm$0.208 & \textbf{0.906$\pm$0.028} & 27.700$\pm$1.485 & 0.848$\pm$0.233 & 1.138$\pm$0.563 \\
~ & $\checkmark$ & $\checkmark$ & $\checkmark$ & $\checkmark$ & \textbf{1.190$\pm$0.227} & 0.900$\pm$0.028 & \textbf{28.118$\pm$1.699} & \textbf{0.698$\pm$0.131} & \textbf{0.922$\pm$0.296} \\
\bottomrule
\end{tabular*}
\end{table*}

\begin{table*}[t]
\centering
\caption{Quantitative comparison of DoseDiff with original mask, TSBDM, ISDM, and PSDM input, respectively.}
\label{table2}
\begin{tabular*}{0.8\textwidth}{@{\extracolsep{\fill}}cccccc}
\toprule
Input & MAE (Gy)$\downarrow$ & SSIM$\uparrow$ & PSNR (dB)$\uparrow$ & Dose score (Gy)$\downarrow$ & Volume score (\%)$\downarrow$ \\
\midrule
Mask & 1.536$\pm$0.244 & 0.859$\pm$0.031 & 25.632$\pm$1.271 & 0.856$\pm$0.238 & \textbf{0.921$\pm$0.372} \\
TSBDM & 1.287$\pm$0.219 & 0.859$\pm$0.029 & 28.099$\pm$1.673 & 0.788$\pm$0.238 & 0.928$\pm$0.304 \\
ISDM & 1.248$\pm$0.214 & 0.878$\pm$0.028 & 28.042$\pm$1.632 & 0.705$\pm$0.181 & 0.928$\pm$0.300 \\
PSDM & \textbf{1.190$\pm$0.227} & \textbf{0.900$\pm$0.028} & \textbf{28.118$\pm$1.699} & \textbf{0.698$\pm$0.131} & 0.922$\pm$0.296 \\
\bottomrule
\end{tabular*}
\end{table*}

\subsection{Evaluation Metrics}
To evaluate the model's performance quantitatively, we used both common image similarity metrics and dosimetry-related metrics. The image similarity metrics included the mean absolute error (MAE), structural similarity index measurement (SSIM), and signal-to-noise ratio (PSNR):

\begin{equation}
\begin{aligned}
\operatorname{MAE}(u, v)=\frac{1}{N} \sum_{i=1}^N\left|u_i-v_i\right|,
\end{aligned}
\end{equation}
\begin{equation}
\begin{aligned}
\operatorname{SSIM}(u, v)=\frac{1}{M}\sum_{j=1}^M \frac{\left(2 \mu_{u,j} \mu_{v,j}+c_1\right)\left(2 \sigma_{uv,j}+c_2\right)}{\left(\mu_{u,j}^2+\mu_{v,j}^2+c_1\right)\left(\sigma_{u,j}^2+\sigma_{v,j}^2+c_2\right)},
\end{aligned}
\end{equation}
\begin{equation}
\begin{aligned}
\operatorname{PSNR}(u, v)=10 \log _{10}\left(\frac{c_{\text {range }}^2}{\frac{1}{N} \sum_{i=1}^N\left(u_i-v_i\right)^2}\right),
\end{aligned}
\end{equation}
where $N$ is the total number of voxels; $u_i$ and $v_i$ are the $i^{th}$ voxel value in volumes $u$ and $v$, respectively. $M$ is the number of local windows in the image. The symbols $\mu_{u,j}$, $\mu_{v,j}$, $\sigma_{u,j}$, $\sigma_{v,j}$, and $\sigma_{uv,j}$ denote local mean, variance, and covariance of the local window $j$ in volumes $u$ and $v$. $c_{range}$ denotes the distance between minimum and maximum possible values of input images, which is set to 3000 in our experiments. Following, \cite{wang2004image}, $c_1=(0.01c_{range})^2$ and $c_2=(0.03c_{range})^2$ are constants to stabilize the division. To eliminate the influence of the background, the mean values over pixels contained within the body masks were computed for these three metrics.

The dosimetry-related metrics included $\triangle D_V$ and $\triangle V_D$. The symbol $\triangle$ denotes the absolute error of the dosimetric metrics between the predicted and ground-truth dose distribution maps. $\triangle D_V$ denotes the dose received by $V\%$ of the volume of the ROIs. $D_{min}$, $D_{mean}$, and $D_{max}$, indicating the minimum, average, and maximum doses in the ROI, respectively, were also considered to be members of $D_V$. For the treatment targets, $V_D$ was the volume percentage of the target region receiving $D\%$ of the prescribed dose; for OARs, $V_D$ was the volume percentage of the ROI receiving $V$-Gy dose. In clinical practice, different indicators are considered for different ROIs. Therefore, we assigned appropriate evaluation metrics for different ROIs in the two datasets. In the breast cancer dataset, we used $\triangle D_{95}$ and $\triangle V_{95}$ for TB and CTV; $\triangle D_{mean}$ and $\triangle V_{30}$ for heart; $\triangle D_{mean}$ and $\triangle V_{20}$ for ipsilateral lung; and $\triangle D_{max}$ for spinal cord. In the NPC dataset, we used $\triangle D_{95}$ and $\triangle V_{95}$ for GTV and PTV; $\triangle D_{max}$ for eyes, optic nerve, temporal lobe, brain stem, and spinal cord; $\triangle D_{mean}$ and $\triangle V_{30}$ for parotid gland; and $\triangle D_{min}$ for mandible. To facilitate model comparison and selection, we defined the mean values of $\triangle D_V$ and $\triangle V_D$ for all ROIs as the “dose score” and “volume score”, respectively. The DVH was used for a comprehensive comparison of the differences between the predicted results obtained with different methods and the real dose distribution.

The dose score and DVH score are employed to evaluate the results of OpenKBP dataset, utilizing evaluation code\footnote{\url{https://github.com/ababier/open-kbp}} directly provided by the event organizers.

\begin{figure}[!t]
\centerline{\includegraphics[width=\columnwidth]{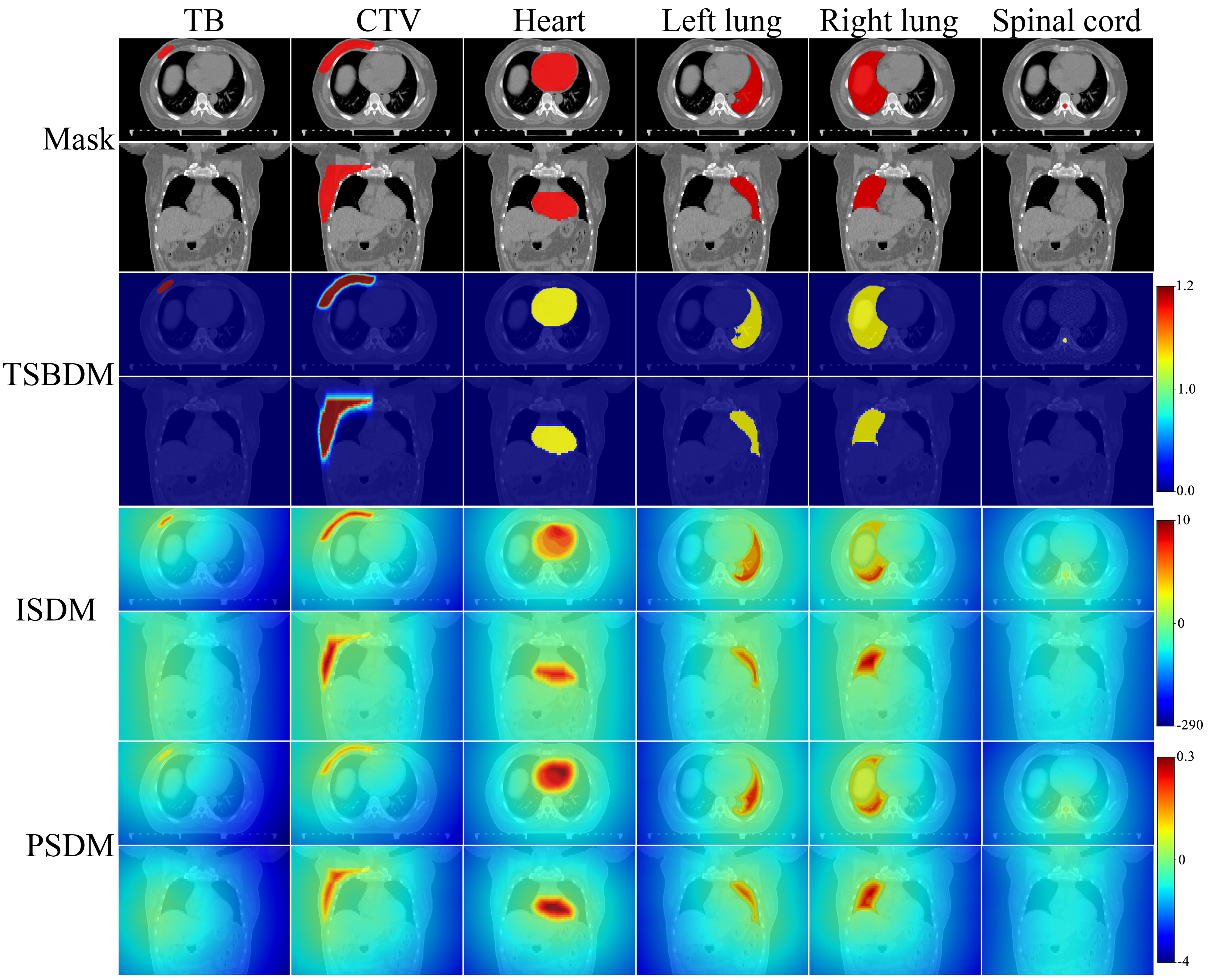}}
\setlength{\abovecaptionskip}{0.05cm}
\caption{Visual comparison of the mask, TSBDM, ISDM, and PSDM with respect to the ROIs of breast cancer.}
\label{Fig3}
\end{figure}

\subsection{Ablation studies for DoseDiff}
We conducted ablation studies to quantitatively validate the contribution of each term in DoseDiff. Specifically, we sequentially added the PSDM input, multi-scale fusion (MS), and fusionFormer module (FF) into the baseline conditional diffusion model, which initially only used CT images as input. Moreover, we added the proposed elements to the baseline UNet model to further prove their effectiveness. As shown in Table \ref{table1}, the introduction of PSDM, MS, and FF individually results in performance improvement, with PSDM contributing the most significant enhancement in performance. Although multi-scale fusion led to limited improvement in the common image similarity metrics, it significantly enhanced the dosimetric metrics. Compared with the conditional diffusion model baseline, DoseDiff led to improvements of 1.747Gy in MAE, 0.093 in SSIM, 7.848dB in RSNR, 9.693Gy in dose score, and 18.491\% in volume score. Furthermore, the experimental results demonstrate that the proposed elements were effective in both the UNet and conditional diffusion model frameworks, with the conditional diffusion model generally outperforming UNet. The results of the ablation experiments indicate that incorporating distance information and enhancing the fusion strategy can be conducive to achieving accurate dose distribution prediction for deep-learning models.

\begin{figure}[!t]
\centerline{\includegraphics[width=1.\columnwidth]{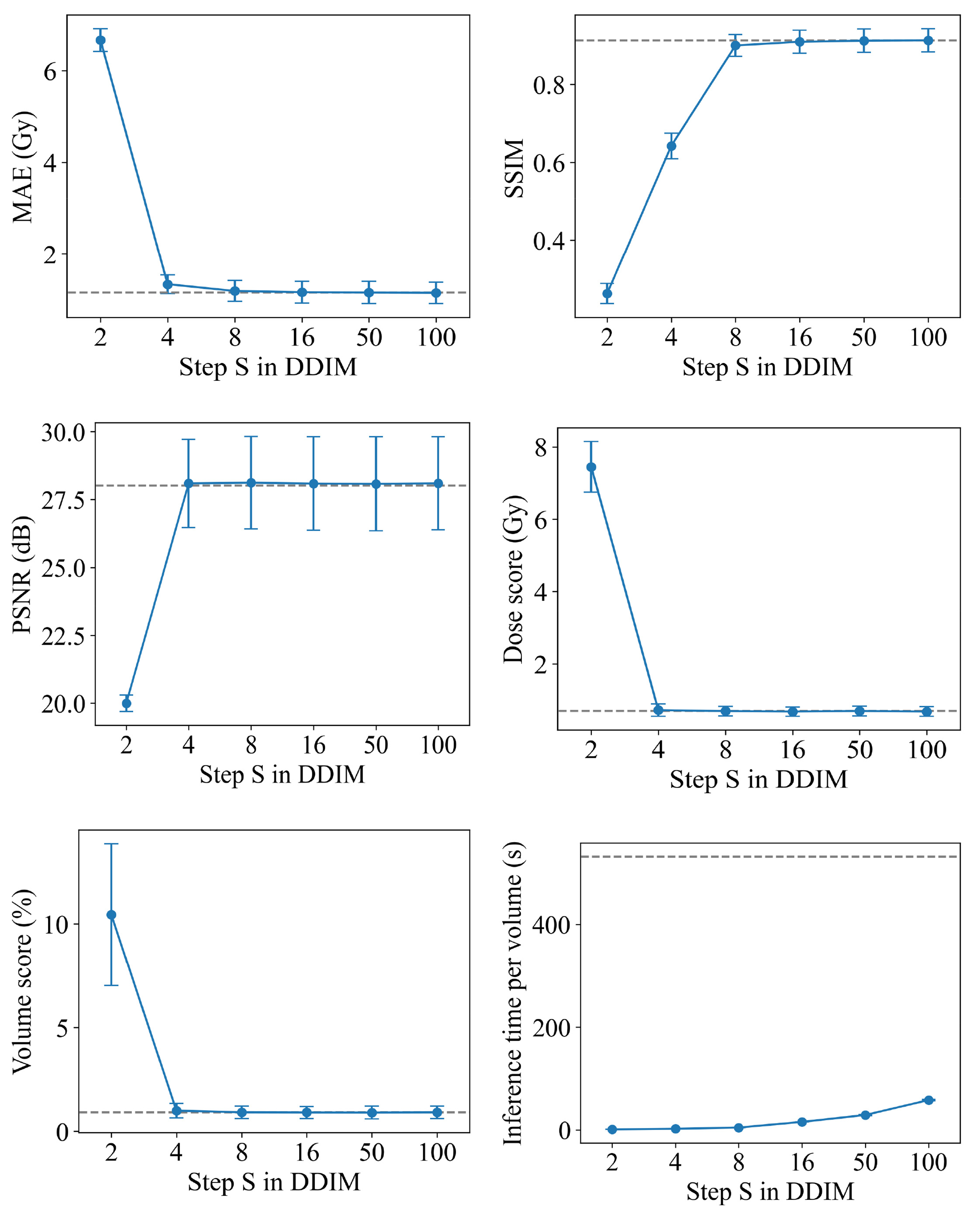}}
\setlength{\abovecaptionskip}{0.05cm}
\caption{Performance and inference time of different generation steps based on DDIM reverse process. Gray dotted lines indicate the metrics of DoseDiff without DDIM ($T=1000$).}
\label{Fig4}
\end{figure}

\subsection{Performance of PSDM}
To verify the benefit of the proposed PSDM, we compared DoseDiff implementations with different distance information inputs, including mask, transformed signed boundary distance map (TSBDM) \cite{yue2022dose}, ISDM, and PSDM. Note that we shrank all ISDM values by a factor of 100 to stabilize the training. Table \ref{table2} shows the quantitative comparison results, demonstration that PSDM achieved the best performance. A visual comparison of the mask, TSBDM, ISDM, and PSDM on the ROIs of breast cancer is presented in Fig. \ref{Fig3}. As shown, the distance maps provided more distinct and effective distance information than the mask images. However, the voxel-wise transformation in TSDBM quickly zeroed out the pixels far from the contour, rendering many background pixels ineffective. In addition, the dynamic change in distance was not obvious in TSDBM. PSDM converts the image distance to the real-world distance, which enabled the model to extract the distance map information in different volumes under a consistent measure. PSDM is also more practical than ISDM in the clinic, for example, ISDM equates the in-plane and inter-slice pixel distance, but the slice thickness of CT images is usually larger than the in-plane spacing.

\begin{figure}[!t]
\centerline{\includegraphics[width=1.\columnwidth]{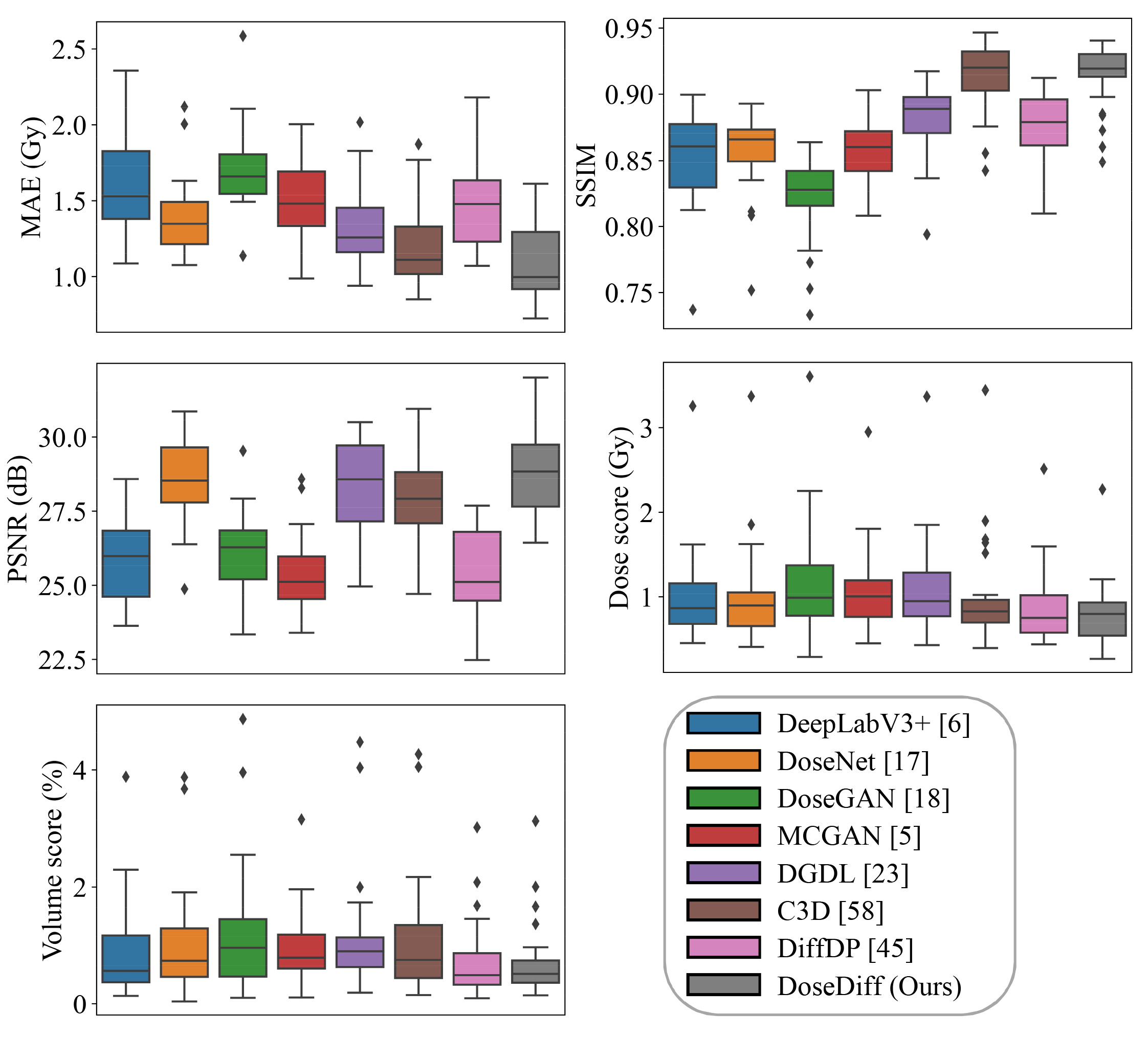}}
\setlength{\abovecaptionskip}{0.05cm}
\caption{Boxplots of dose prediction results using compared methods on the breast cancer dataset.}
\label{Fig4_1}
\end{figure}

\begin{table*}[t]
\centering
\footnotesize
\setlength\tabcolsep{2pt}
\caption{Quantitative comparison with SOTA methods for dose prediction on breast cancer dataset.}
\label{table3}
\begin{tabular}{cccccccccc}
\toprule
ROI & Metrics & DeepLabV3+ & DoseNet & DoseGAN & MCGAN & DGDL & C3D & DiffDP & DoseDiff (Ours) \\
\midrule
\multirow{3}{*}{Body} & MAE (Gy)$\downarrow$ & 1.602$\pm$0.307 & 1.385$\pm$0.256 & 1.706$\pm$0.268 & 1.510$\pm$0.268 & 1.330$\pm$0.242 & 1.195$\pm$0.256 & 1.455$\pm$0.289 & \textbf{1.076$\pm$0.232} \\
~ & SSIM$\uparrow$ & 0.851$\pm$0.034 & 0.857$\pm$0.030 & 0.821$\pm$0.031 & 0.858$\pm$0.026 & 0.880$\pm$0.027 & \textbf{0.913$\pm$0.027} & 0.875$\pm$0.027 & \textbf{0.913$\pm$0.024} \\
~ & PSNR (dB)$\uparrow$ & 25.861$\pm$1.367 & 28.555$\pm$1.452 & 26.121$\pm$1.312 & 25.287$\pm$1.305 & 28.425$\pm$1.376 & 27.981$\pm$1.476 & 25.411$\pm$1.411 & \textbf{28.887$\pm$1.481} \\
\multirow{2}{*}{TB} & $\triangle D_{95}$ (Gy)$\downarrow$ & 0.794$\pm$0.737 & 0.623$\pm$0.416 & 0.854$\pm$0.759 & 1.204$\pm$0.669 & 1.029$\pm$0.657 & \textbf{0.567$\pm$0.404} & 1.183$\pm$0.665 & 1.035$\pm$0.602 \\
~ & $\triangle V_{95}$ (\%)$\downarrow$ & 0.098$\pm$0.254 & \textbf{0.029$\pm$0.074} & 0.503$\pm$0.528 & 0.037$\pm$0.061 & 0.185$\pm$0.327 & 0.031$\pm$0.074 & 0.107$\pm$0.235 & 0.084$\pm$0.194 \\
\multirow{2}{*}{CTV} & $\triangle D_{95}$ (Gy)$\downarrow$ & 0.713$\pm$0.387 & 1.115$\pm$0.420 & 0.620$\pm$0.299 & 0.844$\pm$0.332 & 1.038$\pm$0.385 & 0.975$\pm$0.404 & 0.378$\pm$0.313 & \textbf{0.328$\pm$0.261} \\
~ & $\triangle V_{95}$ (\%)$\downarrow$ & 0.377$\pm$0.301 & 0.562$\pm$0.463 & 0.495$\pm$0.448 & 0.593$\pm$0.462 & 0.582$\pm$0.456 & 0.557$\pm$0.466 & 0.290$\pm$0.411 & \textbf{0.201$\pm$0.214} \\
\multirow{2}{*}{Heart} & $\triangle D_{mean}$ (Gy)$\downarrow$ & 1.289$\pm$1.408 & 1.120$\pm$1.454 & 1.503$\pm$1.802 & 0.996$\pm$1.330 & 1.293$\pm$1.596 & 1.191$\pm$1.537 & 1.044$\pm$1.179 & \textbf{0.938$\pm$1.061} \\
~ & $\triangle V_{30}$ (\%)$\downarrow$ & 1.679$\pm$2.541 & 2.021$\pm$3.549 & 2.094$\pm$3.775 & 1.081$\pm$1.810 & 2.026$\pm$3.645 & 2.231$\pm$3.819 & 1.337$\pm$2.091 & \textbf{1.315$\pm$2.026} \\
\multirow{2}{1cm}{\centering Ipsilateral\\ lung} & $\triangle D_{mean}$ (Gy)$\downarrow$ & 1.179$\pm$0.986 & 0.910$\pm$0.704 & 1.491$\pm$1.069 & 1.071$\pm$0.787 & 1.105$\pm$0.927 & 1.011$\pm$0.741 & \textbf{0.818$\pm$0.717} & 0.872$\pm$0.737 \\
~ & $\triangle V_{20}$ (\%)$\downarrow$ & 1.566$\pm$1.465 & 1.709$\pm$1.269 & 1.772$\pm$1.695 & 2.092$\pm$1.325 & 1.868$\pm$1.246 & 1.508$\pm$1.185 & \textbf{1.229$\pm$1.034} & 1.345$\pm$1.021 \\
Spinal cord & $\triangle D_{max}$ (Gy)$\downarrow$ & 1.037$\pm$1.050 & 1.376$\pm$1.317 & 1.557$\pm$1.273 & 1.183$\pm$1.018 & 1.238$\pm$1.284 & 1.308$\pm$1.307 & 1.044$\pm$1.016 & \textbf{0.877$\pm$0.812} \\
\bottomrule
\end{tabular}
\end{table*}

\begin{figure*}[!t]
\centerline{\includegraphics[width=1.\textwidth]{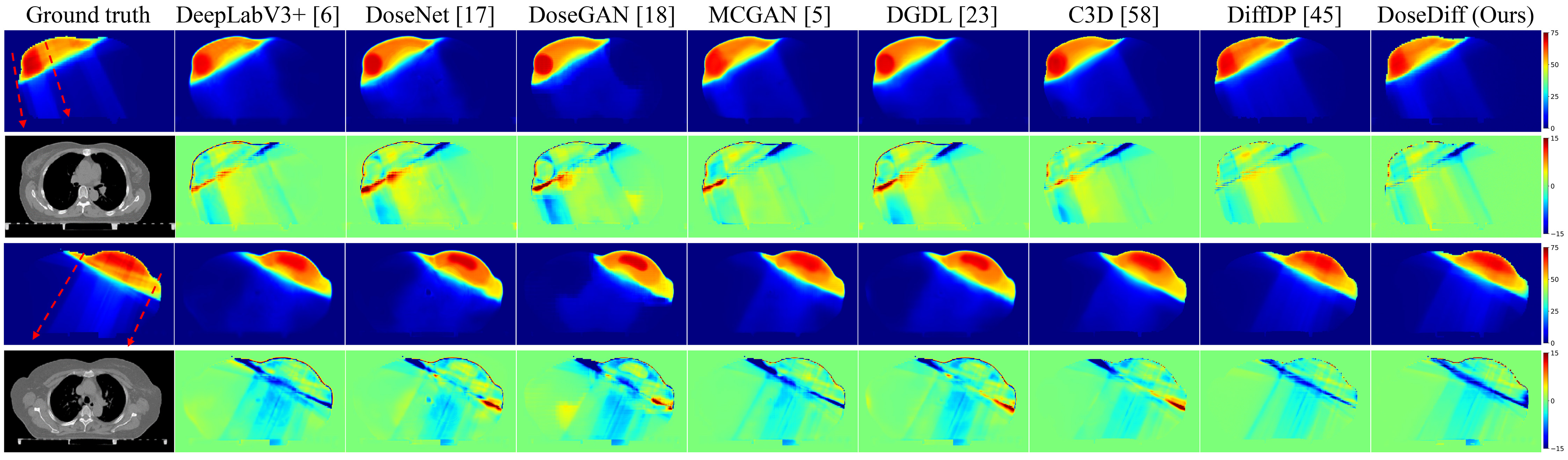}}
\setlength{\abovecaptionskip}{0.05cm}
\caption{Visual comparison of predicted dose distribution maps obtained with different methods on the breast cancer dataset. Red dashed arrows represent the estimated input directions of certain rays identified from the real dose distribution maps.}
\label{Fig5}
\end{figure*}

\begin{figure*}[!t]
\centerline{\includegraphics[width=1.\textwidth]{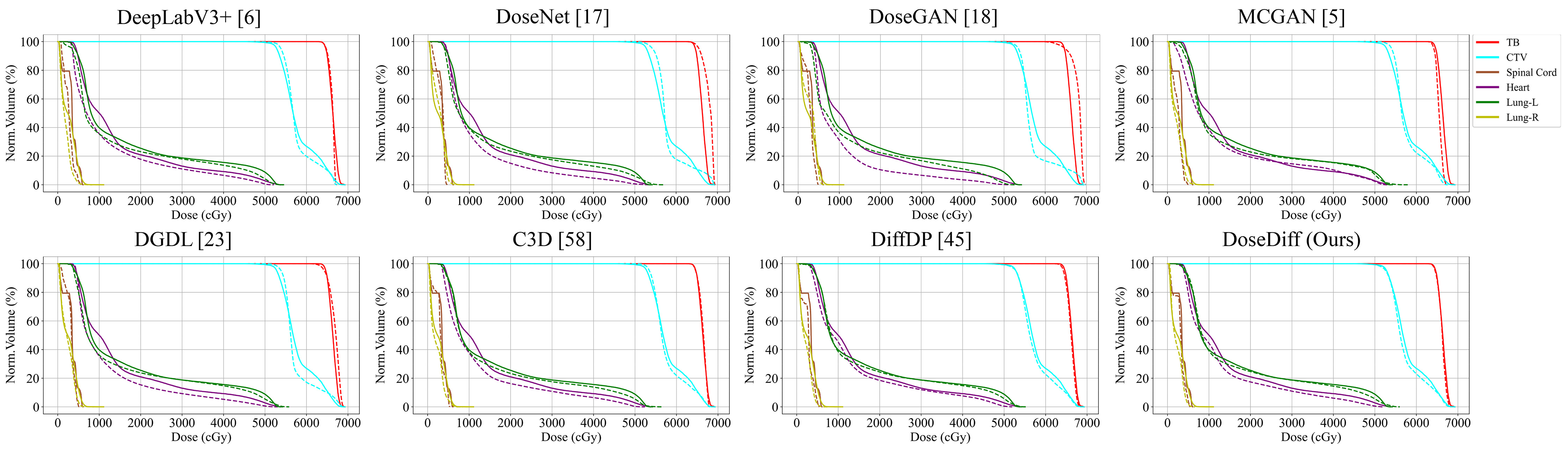}}
\setlength{\abovecaptionskip}{0.05cm}
\caption{DVHs of the dose distribution maps predicted by different methods for a breast cancer patient in the test set. The solid line represents the ground-truth DVH; dotted line represents the DVH of the predicted dose distribution map.}
\label{Fig6}
\end{figure*}

\subsection{Trade-off between performance and inference speed}
\label{DDIM}
In diffusion models, the generative process is defined as the reverse of a particular Markovian diffusion process. Consequently, the generation of the image necessitates a complete sequence of sampling, which is time-consuming. The DDIM converts the Markovian diffusion process in the diffusion model into a non-Markovian diffusion process that allows us to trade computation for sample quality. Based on the DDIM, we conducted a comparison of the performance and inference time of DoseDiff across varying generation steps. As shown in Fig. \ref{Fig4}, the performance of the model improved with more generation steps but stabilized after eight steps. The inference time was approximately proportional to the number of generation steps. To trade off performance and inference time, we suggest setting the generation step to 8 in the DDIM reverse process, resulting in an inference time of 4.7s per volume; our subsequent experiments all adopt this setting.

\begin{table*}[t]
\centering
\footnotesize
\setlength\tabcolsep{2pt}
\caption{Quantitative comparison with SOTA methods for dose prediction on NPC dataset.}
\label{QC_NPC}
\begin{tabular}{cccccccccc}
\toprule
ROI & Metrics & DeepLabV3+ & DoseNet & DoseGAN & MCGAN & DGDL & C3D & DiffDP & DoseDiff (Ours) \\
\midrule
\multirow{3}{*}{Body} & MAE (Gy)$\downarrow$ & 2.075$\pm$0.309 & 1.923$\pm$0.346 & 2.495$\pm$0.435 & 1.973$\pm$0.359 & 1.926$\pm$0.346 & 1.851$\pm$0.378 & 1.965$\pm$0.424 & \textbf{1.676$\pm$0.387} \\
~ & SSIM$\uparrow$ & 0.827$\pm$0.042 & 0.839$\pm$0.034 & 0.776$\pm$0.082 & 0.856$\pm$0.030 & 0.837$\pm$0.044 & 0.860$\pm$0.031 & 0.867$\pm$0.037 & \textbf{0.905$\pm$0.018} \\
~ & PSNR (dB)$\uparrow$ & 25.880$\pm$1.048 & \textbf{26.430$\pm$1.304} & 24.592$\pm$1.044 & 25.580$\pm$1.071 & 26.445$\pm$1.191 & 26.256$\pm$1.341 & 25.526$\pm$1.254 & 26.397$\pm$1.249 \\
\multirow{2}{*}{GTV} & $\triangle D_{95}$ (Gy)$\downarrow$ & 0.758$\pm$0.585 & \textbf{0.596$\pm$0.475} & 0.858$\pm$0.684 & 0.756$\pm$0.518 & 0.637$\pm$0.374 & 1.831$\pm$0.553 & 1.777$\pm$1.282 & 1.243$\pm$0.677 \\
~ & $\triangle V_{95}$ (\%)$\downarrow$ & 0.494$\pm$0.488 & 0.500$\pm$0.548 & 0.889$\pm$0.975 & \textbf{0.430$\pm$0.560} & 0.516$\pm$0.504 & 0.585$\pm$0.691 & 0.996$\pm$1.177 & 0.697$\pm$0.842 \\
\multirow{2}{*}{PTV} & $\triangle D_{95}$ (Gy)$\downarrow$ & 8.695$\pm$9.358 & 7.896$\pm$7.309 & 7.822$\pm$9.289 & 3.072$\pm$3.389 & 7.475$\pm$7.005 & 6.666$\pm$6.334 & 2.977$\pm$2.220 & \textbf{2.623$\pm$2.322} \\
~ & $\triangle V_{95}$ (\%)$\downarrow$ & 4.285$\pm$3.283 & 3.740$\pm$2.199 & 6.595$\pm$4.172 & 5.534$\pm$3.660 & 3.669$\pm$2.282 & \textbf{2.441$\pm$1.84} & 5.032$\pm$3.284 & 4.247$\pm$2.811 \\
\multirow{2}{*}{parotid glan} & $\triangle D_{mean}$ (Gy)$\downarrow$ & 2.229$\pm$1.492 & 1.804$\pm$1.497 & 2.312$\pm$1.196 & 1.581$\pm$1.237 & 1.962$\pm$1.361 & \textbf{1.538$\pm$1.006} & 2.144$\pm$1.308 & 1.944$\pm$1.118 \\
~ & $\triangle V_{30}$ (\%)$\downarrow$ & 8.684$\pm$5.510 & 5.898$\pm$4.276 & 4.814$\pm$3.722 & \textbf{4.808$\pm$3.878} & 5.736$\pm$3.420 & 6.779$\pm$5.191 & 5.404$\pm$3.640 & 5.112$\pm$3.300 \\
Eyes & $\triangle D_{max}$ (Gy)$\downarrow$ & 6.106$\pm$5.561 & 5.801$\pm$5.836 & 5.601$\pm$3.816 & 6.442$\pm$6.387 & 6.163$\pm$4.553 & \textbf{5.087$\pm$4.245} & 7.926$\pm$4.247 & 6.449$\pm$4.271 \\
Optic nerve & $\triangle D_{max}$ (Gy)$\downarrow$ & 5.753$\pm$4.525 & 2.808$\pm$2.637 & 2.947$\pm$3.031 & 6.164$\pm$6.851 & 4.244$\pm$3.479 & \textbf{2.871$\pm$2.976} & 7.879$\pm$7.959 & 6.324$\pm$6.922 \\
Temporal lobe & $\triangle D_{max}$ (Gy)$\downarrow$ & 3.130$\pm$1.593 & 4.446$\pm$2.241 & 3.739$\pm$2.863 & \textbf{2.659$\pm$1.615} & 3.743$\pm$2.528 & 3.361$\pm$2.530 & 3.948$\pm$2.849 & 4.165$\pm$2.531 \\
Brain stem & $\triangle D_{max}$ (Gy)$\downarrow$ & 4.030$\pm$1.863 & 4.033$\pm$1.770 & 4.960$\pm$1.865 & 3.842$\pm$1.675 & 4.334$\pm$1.927 & 3.623$\pm$1.810 & 4.118$\pm$2.649 & \textbf{3.616$\pm$2.276} \\
Mandible & $\triangle D_{min}$ (Gy)$\downarrow$ & 7.206$\pm$5.779 & 7.006$\pm$6.568 & 5.466$\pm$5.957 & 5.226$\pm$5.801 & 8.627$\pm$6.366 & 6.639$\pm$5.807 & 4.674$\pm$4.500 & \textbf{4.483$\pm$4.798} \\
Spinal cord & $\triangle D_{max}$ (Gy)$\downarrow$ & 2.760$\pm$1.543 & 1.602$\pm$0.914 & \textbf{1.421$\pm$0.913} & 4.855$\pm$0.868 & 2.405$\pm$0.888 & 2.558$\pm$1.418 & 2.302$\pm$1.252 & 1.992$\pm$1.293 \\
\bottomrule
\end{tabular}
\end{table*}

\begin{figure}[!t]
\centerline{\includegraphics[width=1.\columnwidth]{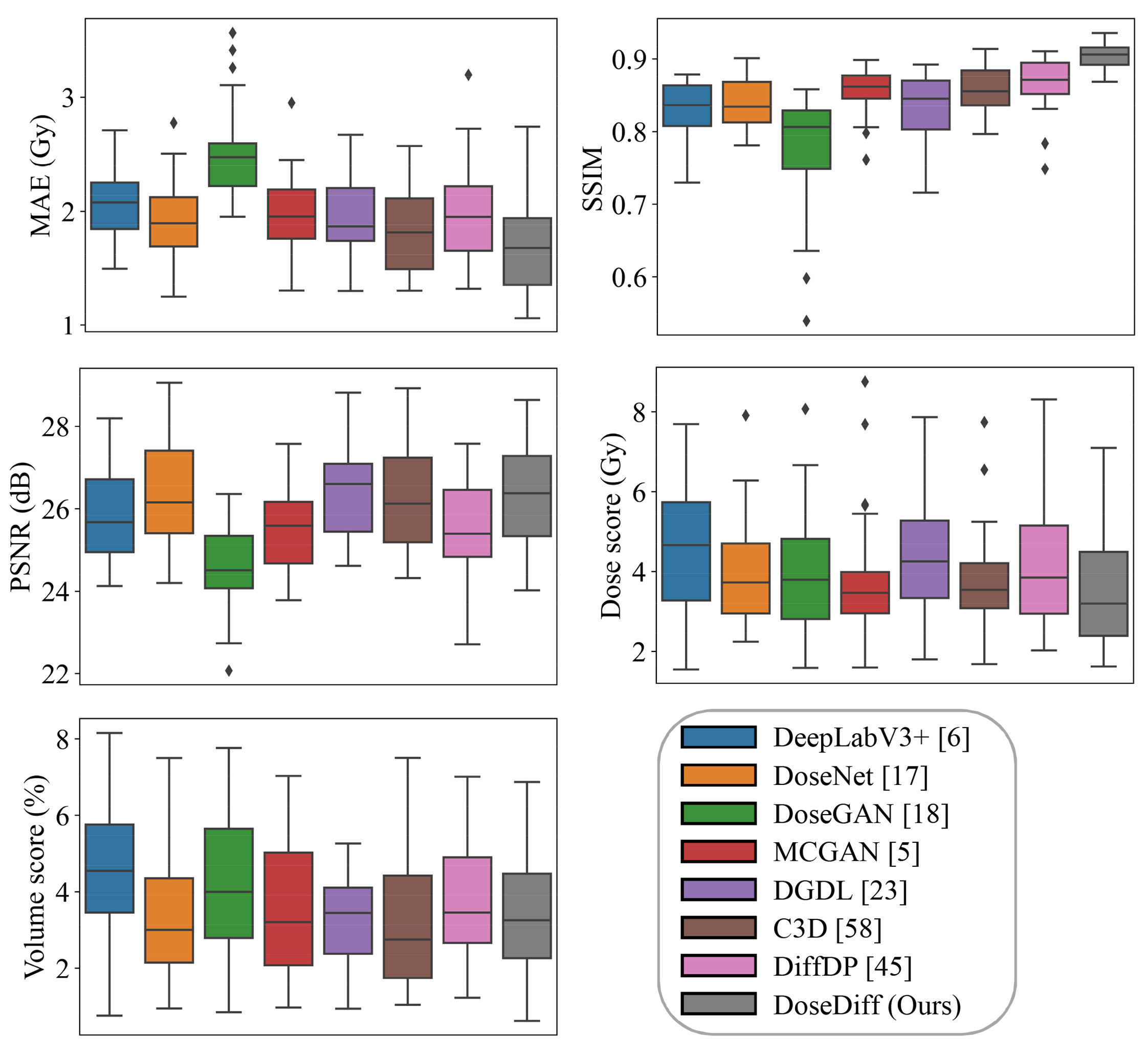}}
\setlength{\abovecaptionskip}{0.05cm}
\caption{Boxplots of dose prediction results using compared methods on the NPC dataset.}
\label{Fig7}
\end{figure}

\subsection{Comparison with SOTA methods}
To demonstrate the superiority of our method, we conducted a comparison between DoseDiff and several SOTA dose prediction methods, including the 2D models DeeplapV3+ \cite{song2020dose}, MCGAN \cite{zhan2022multi}, and DiffDP \cite{feng2023diffdp}, and the 3D models DoseNet \cite{kearney2018dosenet}, DoseGAN \cite{kearney2020dosegan}, distance-guided deep learning (DGDL) \cite{yue2022dose}, and cascade 3D U-Net (C3D) \cite{liu2021cascade}. DGDL differs from the other methods in its use of TSBDM distance maps as input; the other methods use mask images. We conducted both quantitative and visual comparisons on the breast cancer and NPC datasets.

\textbf{\textit{Comparison on breast cancer dataset:}} Table \ref{table3} and Figure \ref{Fig4_1} illustrate the common image similarity metrics and detailed dosimetry-related metrics for all methods compared on the breast cancer dataset. As shown in the Table \ref{table3}, our method achieved stable top performance on most evaluation metrics. We can observe that although the C3D and DoseNet exhibit excellent predictive performance on the TB, their performance on other ROIs is relatively poor. We speculate that they may be overfitting to the dose distribution in the TB, which is a region with relatively uniform and fixed cumulative dose values (e.g., 63.8 Gy for our dataset). Figure \ref{Fig5} shows predicted dose distribution maps obtained with different methods, as well as the difference between the dose distribution maps and the ground truths. The dose distribution predicted by our method was closest closer to the real dose distribution. Specifically, our method can learn a clear path of the light ray that reflects the fundamental properties of straight-line propagation, which is beneficial for medical physicists to obtain precise parameters, such as the number of radiation fields and the incident angle, from the predicted dose distribution map. Furthermore, Fig. \ref{Fig6} presents the DVHs of the dose distribution maps predicted by the compared methods, allowing for a comprehensive comparison of the dose distributions. We also observed that the DVH curve of the dose distribution map predicted by our method was most similar to the ground-truth curve.

\begin{table}[t]
\centering
\caption{Quantitative comparison with SOTA methods for dose prediction on OpenKBP dataset.}
\label{QC_OpenKBP}
\begin{tabular*}{\columnwidth}{@{\extracolsep{\fill}}lcc}
\toprule
Methods & Dose score $\downarrow$ & DVH score $\downarrow$ \\
\midrule
DeepLabV3+ \cite{song2020dose} & 3.551$\pm$1.126 & 2.081$\pm$2.169 \\
DoseNet \cite{kearney2018dosenet} & 2.861$\pm$1.107 & 1.647$\pm$1.926 \\
DoseGAN \cite{kearney2020dosegan} & 2.873$\pm$1.125 & 1.695$\pm$1.973 \\
MCGAN \cite{zhan2022multi} & 3.442$\pm$1.081 & 1.796$\pm$2.210 \\
DGDL \cite{yue2022dose} & 3.026$\pm$1.085 & 1.748$\pm$2.033 \\
C3D \cite{liu2021cascade} & 2.429$\pm$1.101 & 1.478$\pm$1.931 \\
DiffDP \cite{feng2023diffdp} & 3.173$\pm$0.966 & 1.602$\pm$1.889 \\
DoseDiff (Ours) & \textbf{2.382$\pm$0.925} & \textbf{1.476$\pm$1.645} \\
\bottomrule
\end{tabular*}
\end{table}

\begin{figure*}[!t]
\centerline{\includegraphics[width=1.\textwidth]{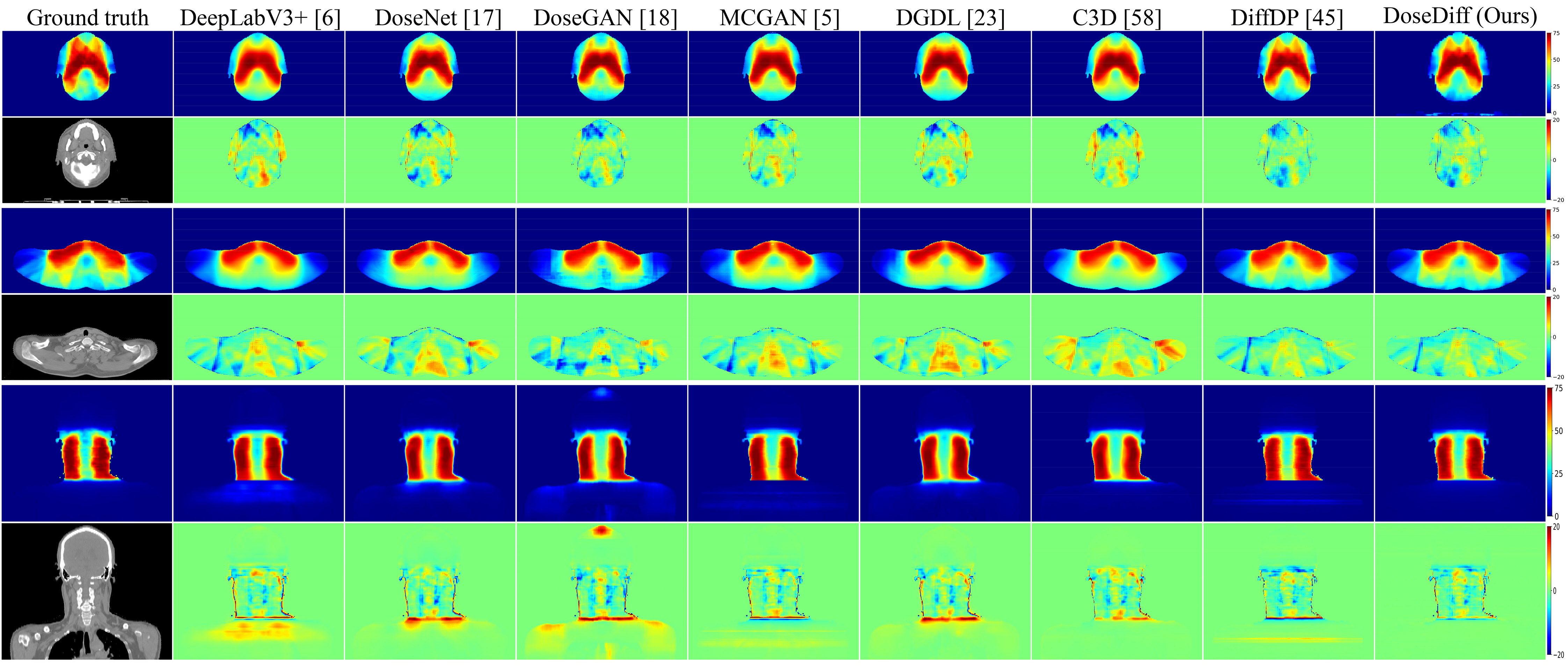}}
\setlength{\abovecaptionskip}{0.05cm}
\caption{Visual comparison of predicted dose distribution maps obtained with different methods on the NPC dataset.}
\label{Fig8}
\end{figure*}

\begin{figure*}[!t]
\centerline{\includegraphics[width=\textwidth]{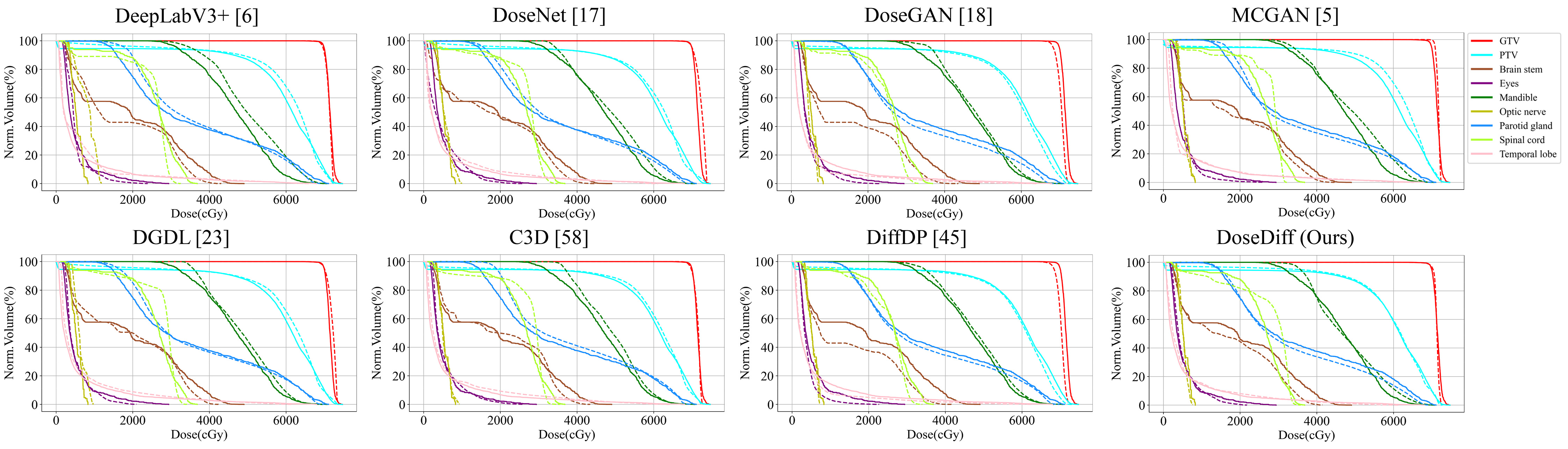}}
\setlength{\abovecaptionskip}{0.05cm}
\caption{The dose-volume histograms (DVHs) of the dose distribution maps predicted by the different methods on a testing NPC patient. The solid line represents the ground-truth DVH, while the dotted line represents the DVH of the predicted dose distribution map.}
\label{Fig9}
\end{figure*}

\textbf{\textit{Comparison on NPC dataset:}} Table \ref{QC_NPC} and Figure \ref{Fig7} present the results of a quantitative comparison between our method and the SOTA approaches on the NPC dataset. Our method achieved competitive results in terms of both image similarity and dosimetry-related metrics. The transverse and coronal plane samples of dose distribution maps predicted by all the compared methods are shown in Fig. \ref{Fig8}. Consistent with the performance observed on the breast cancer dataset, the dose distribution maps predicted by previous methods were overly smooth, whereas our method's predictions retained more realistic ray path characteristics. Furthermore, we could observe from the coronal images that our method maintained a high level of prediction accuracy for regions that were far from the ROIs, owing to the distance information provided by the PSDM. The DVH curves of the dose distribution maps predicted by the compared methods on the NPC dataset are presented in Fig. \ref{Fig9}. Compared to other methods, the DVH curves of the target and OARs predicted by our method are closer to the ground truth.

\begin{table*}[!t]
\centering
\caption{The variance of each mean metric for our DoseDiff with 50 different Gaussian samplings on the two datasets.}
\label{table4}
\begin{tabular*}{0.7\textwidth}{@{\extracolsep{\fill}}cccccc}
\toprule
Dataset & MAE (Gy) & SSIM & PSNR (dB) & Dose score (Gy) & Volume score (\%) \\
\midrule
Breast cancer & 0.0011 & 0.0002 & 0.0099 & 0.0058 & 0.0029 \\
NPC & 0.0012 & 0.0001 & 0.0076 & 0.0226 & 0.0153 \\
\bottomrule
\end{tabular*}
\end{table*}

\begin{figure*}[!t]
\centerline{\includegraphics[width=1.\textwidth]{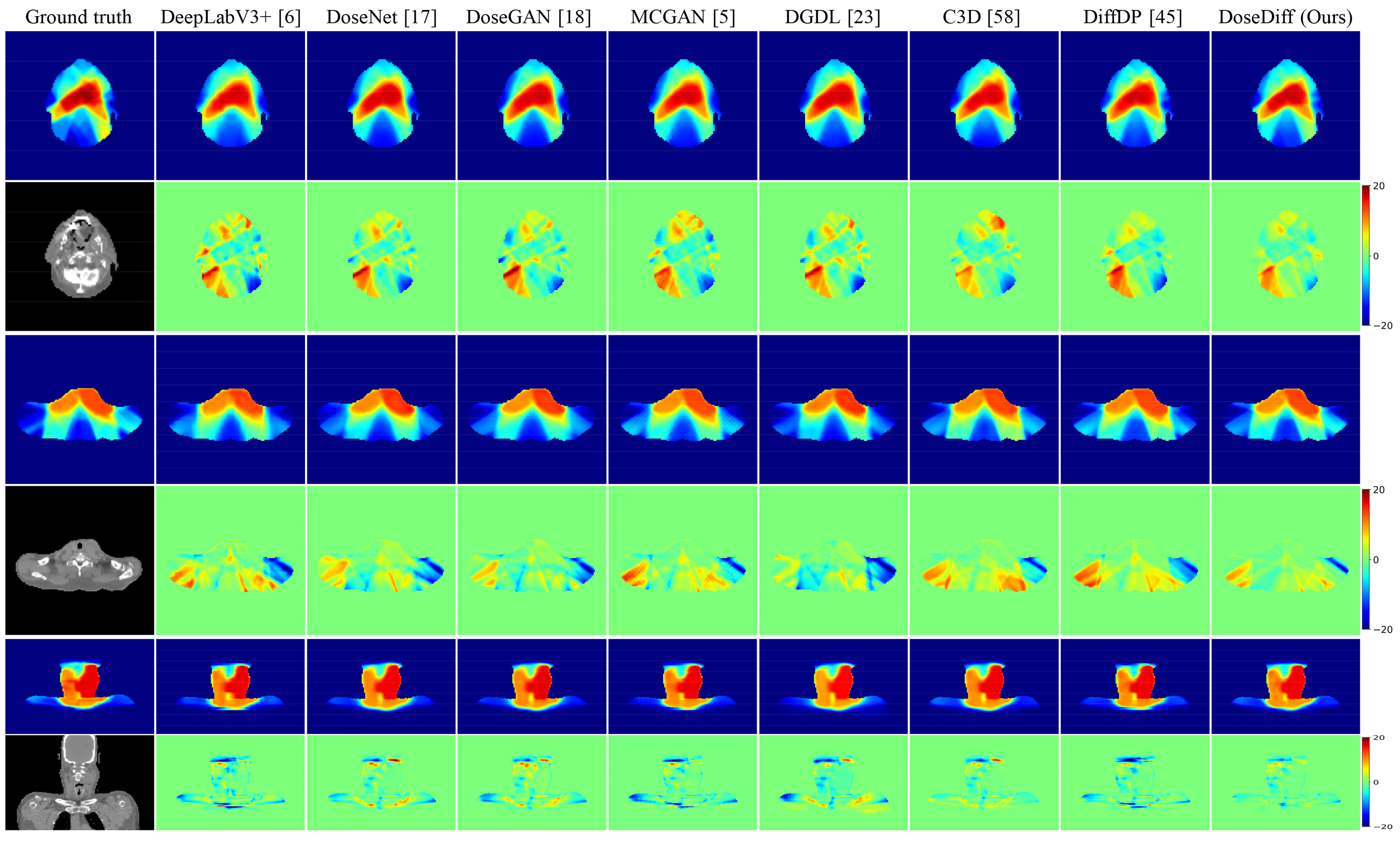}}
\setlength{\abovecaptionskip}{0.05cm}
\caption{Visual comparison of predicted dose distribution maps obtained with different methods on the OpenKBP dataset.}
\label{VC_OpenKBP}
\end{figure*}

\textbf{\textit{Comparison on OpenKBP dataset:}} The results of a quantitative comparison between our method and the SOTA methods on the OpenKBP dataset are shown in Table \ref{QC_OpenKBP}. The images in the OpenKBP dataset exhibit larger in-plane spacings and smaller slice thickness (compared to internal datasets), implying that dose prediction on this dataset relies more heavily on 3D information. Consequently, we can observe that 2D models generally underperform compared to 3D models on the OpenKBP dataset. Despite also being a 2D model, the introduction of 3D distance information by PSDM in our DoseDiff enables our method to achieve competitive results. Figure \ref{VC_OpenKBP} shows the transverse and coronal plane samples of dose distribution maps predicted by all the compared methods. The visualization results demonstrate that our method exhibits minimal error between predicted values and ground truth.

\subsection{The influence of non-deterministic prediction}
Each dose prediction in our DoseDiff begins with a randomly generated Gaussian noise image. Moreover, each step in the DDIM-based inverse process involves Gaussian sampling as in Eqs. \ref{eqs7} and \ref{eqs8}. The randomness introduced by these two factors leads to the non-determinism in dose prediction of our model. To investigate the influence of this non-determinism on the prediction results, we conducted 50 dose predictions for both breast cancer and NPC datasets using different random seeds. The variance of each mean metric for our DoseDiff on the two datasets are shown in Table \ref{table4}, and the results indicate a small degree of fluctuation for each metric.

\section{Discussion}
\textcolor{black}{In this study, we aimed to develop a model to accurately predict dose distribution for individual RT patients, enabling medical physicists to achieve acceptable RT plans with less trial and error. Clinically, both CT images and ROI masks are essential for RT planning, so most previous studies have used deep learning to directly learn mappings from them to dose distribution maps. However, the mask image prevents the deep learning model from effectively extracting distance information between surrounding tissues and targets or OARs. Some studies have replaced masks with distance maps and observed improved dose prediction accuracy as a result \cite{kontaxis2020deepdose,yue2022dose}, because each voxel of the distance map explicitly provides distance information relative to the ROI contour. Here,} inspired by the success of diffusion models, we propose a distance-aware conditional diffusion model called DoseDiff for dose prediction. Using distance maps as input, we also explore a fusion strategy involving distance maps and CT images. We incorporate multi-scale fusion strategy and fusionFormer module in the proposed MMFNet to achieve effective fusion of complex information

Diffusion models naturally possess a strong ability to model data distributions, as they were specifically developed for this purpose. Therefore, one advantage of introducing conditional diffusion model into image-to-image translation tasks is that as well as learning the mapping relationship from conditions to output, it can also effectively capture the distribution characteristics of the generated image itself. By contrast, previous methods have only learned the mapping relationship between input and output images. As shown in Figs. \ref{Fig5} and \ref{Fig8}, that the dose distribution maps generated by our method are more realistic than those obtained with previous methods, in which the ray path characteristics are unclear and distorted. These dose distribution maps may not exist in the real world and may be limited in their usefulness for assisting medical physicists in obtaining the relevant parameters for treatment planning.

Unlike the previous single-step methods, our conditional diffusion model predicts dose distribution through a multi-step denoising process. This paradigm progressively recovers prediction results from Gaussian noise in a recursive manner, which improves the reliability and realism of the results. As shown in Table \ref{table1}, the conditional diffusion model outperformed the single-step UNet model in various metrics on dose prediction. However, the multi-step paradigm inevitably increases the time for model training and inference. We explore the relationship between the performance and inference time of DoseDiff in different generating steps in Section \ref{DDIM}. Our experimental results demonstrate that DoseDiff, using DDIM technology, is able to generate an accurate dose distribution map for a new patient in only eight steps.

Proper use of distance maps can improve the performance of dose prediction models. The unit distance in the vanilla SDM (i.e., ISDM) is one voxel, so the magnitude of the values in ISDM are usually comparable to the volume size. Such large values are usually not conducive to the stable training of neural networks. Therefore, Yue et al. \cite{yue2022dose} proposed a voxel-wise transformation to shrink the values in SDM. The transformation keeps values inside the ROI large, while gradually decreasing the values outside the ROI to eventually approach zero; this makes the numerical distribution trend in SDM similar to that of the dose distribution. However, this transformation is too drastic, as it causes large numbers of voxels outside the ROI to become zero, rendering them unable to provide effective distance information. Our experiments demonstrate that the ISDM with simple division by a normalization constant achieves better performance than TSBDM, as shown in Table \ref{table2}. The PSDM can be regarded as a global normalized version of ISDM with the introduction of spacing information. The transformation in TSBDM is still the instance-wise normalization, whereas our proposed PSDM normalizes the values in all SDM to a unified distance unit (e.g., a decimeter). Therefore, the model can extract and analyze the information in PSDM more efficiently. In addition, PSDM is calculated in 3D volume, so every voxel in PSDM contains some 3D information. Although DoseDiff is a 2D model, with the support of PSDM, its performance is not inferior to that of SOTA 3D models (i.e., DoseNet, DoseGAN, and DGDL).

Existing dose prediction methods usually adopt early fusion for CT image and mask/SDM, i.e., concatenating them at the input level. This fusion strategy is easy to implement, but it is often suboptimal. Therefore, we propose MMFNet to extract more valuable interactive information between the CT image and PSDM. MMFNet uses multi-scale feature-level fusion for CT image and PSDM. Moreover, a fusionFormer module is used to perceive long-range dependence and global information fusion. Long-range perception is important for dose prediction because the dose of a voxel is related to all the tissues through which the ray passes before reaching that voxel. Transformers have better long-range perception capability compared with CNNs, so we designed a transformer-based fusion module. The experimental results in Table 1 show the effectiveness of our fusion strategy.

Our work had several limitations. First, the 3D information provided by PSDM is relatively limited, and our method is unable to use 3D structural information from CT images, which would be beneficial to dose prediction. Therefore, our future work will focus on expanding DoseDiff into a 3D model. Second, DoseDiff has no advantage in inference speed compared with one-step dose prediction models. The diffusion model paradigm inevitably incurs more inference costs. Fortunately, approaches such as DDIM have been proposed to reduce the inference time of the diffusion model and preserve the model's performance. Third, while non-deterministic predictions have limited influence on model performance as shown in Table \ref{table4}, they remain an unresolved issue that requires attention. This is a common challenge for applying conditional diffusion model to image-to-image tasks.

\section{Conclusion}
In this paper, we propose DoseDiff, a distance-aware conditional diffusion model, for predicting dose distribution maps. DoseDiff uses CT images and SDMs as conditions and dose distribution map prediction is defined as a sequence of denoising steps guided by these conditions. Using PSDM as a model input results in better performance compared with using a mask, as it can effectively provide distance information. Moreover, MMFNet is proposed to effectively extract and fuse features from CT images and SDMs. We used ablation studies to evaluate the contribution of each element of DoseDiff. Comparison with other methods on two datasets showed that DoseDiff achieves SOTA performance in dose prediction.

\bibliographystyle{IEEEtran}
\bibliography{IEEEexample}

\end{document}